%% file: XeDopedArMaster.tex
\documentclass[preprint,12pt]{elsarticle}

\newcommand{\XeD}{XeDLAr}  
\newcommand{\BBz}       {$0\nu\beta\beta$}
\newcommand{\singlet} {$^{1}\Sigma^{+}_{u}$ }
\newcommand{\triplet} {$^{3}\Sigma^{+}_{u}$ }

\usepackage{amssymb}

\usepackage{graphicx}
\usepackage{caption}
\usepackage{amsmath}
\usepackage{subcaption}
\usepackage{mathtools}
\usepackage[version=3]{mhchem}

\usepackage{physics}
\usepackage{dirtytalk} 
\usepackage{xfrac}
\usepackage{textcomp}
\usepackage{placeins}
\usepackage{ulem}

\usepackage[compact]{titlesec}

\usepackage{float}

\usepackage{etoolbox}

\usepackage[page,toc,titletoc,title]{appendix}
\usepackage{tocloft}

\usepackage{blindtext}
\usepackage{xcolor}
\journal{Nuclear Instruments and Methods A}

\begin{document}

\begin{frontmatter}



\title{Large-Scale, Precision Xenon Doping of Liquid Argon}


\author[UNM]{N.~McFadden}
\author[LANL]{S.~R.~Elliott}
\author[UNM]{M.~Gold}
\author[UNM]{D.E.~Fields}
\author[LANL]{K.~Rielage}
\author[LANL]{R.~Massarczyk}
\author[UNM]{R.~Gibbons }

\address[UNM]{Department of Physics and Astronomy
MSC07 4220, 1 University of New Mexico,Albuquerque NM 87131-0001}
\address[LANL]{Physics Division, Los Alamos National Laboratory MS H803, P-23, Los Alamos, NM, 87545, USA}


\begin{abstract}
\input{abtract.tex}
\end{abstract}




\end{frontmatter}


\input{0_motivation.tex}
\input{1_ArgonandXenonDopingScintillationFeatures.tex}
\input{2_experimental.tex}

\input{3_Triplet-Lifetime.tex}
\input{4_NitrogenDoping.tex}
\input{6_LightYeildComparison.tex}
\input{7_conclusion.tex}
\input{8_Acknowledgements.tex}

\bibliography{mybibfile}{}                    
\bibliographystyle{acm}








\end{document}

%% file: abtract.tex
\label{sec abstract}

The detection of scintillation light from liquid argon is an experimental technique key to a number of current and future nuclear/particle physics experiments, such as neutrino physics, neutrinoless double beta decay and dark matter searches. Although the idea of adding small quantities of xenon (doping) to enhance the light yield has attracted considerable interest, this technique has never been demonstrated at the necessary scale or precision.  Here we report on xenon doping in a 100~l cryogenic vessel. Xenon doping was performed in four concentrations of 1.00$\pm$0.06~ppm, 2.0$\pm$0.1~ppm, 5.0$\pm$0.3~ppm, and 10.0$\pm$0.5~ppm. These measurements represent the most precise xenon doping measurements as of publishing. We observed an increase in average light yield by a factor of 1.92$\pm$0.12(syst)$\pm$0.02(stat) at a dopant concentration of 10~ppm.

%% file: 0_motivation.tex
\section{Introduction}
\label{sec Intro}
Liquid noble-gas detectors (LNGD) are widely used or are planned to be used in many different particle physics applications, including dark matter (DARWIN \cite{Aalbers:2016jon}), neutrino tracking (DUNE \cite{DUNE}), and neutrinoless double beta decay (\BBz) (GERDA~\cite{Agostini:2015boa,Agostini:2018}, LEGEND \cite{LEGENDOverview}).
Although He and Ne have been considered~\cite{LNGD_LE}, Ar and Xe monolithic LNGDs have advantages due to the large scintillation yields, longer scintillation wavelengths, and fast scintillation time constants. For the liquid Ar (LAr) case, the boiling temperature is near the preferred operating temperature of Ge detectors and Ar can provide substantial radiation shielding. These features led to the GERDA concept of operating Ge detectors bare within a LAr environment~\cite{Agostini:2015boa}. 
If the light detection of the LAr in GERDA could be significantly enhanced, it would result in improved background rejection. Doping the LAr with Xe is one technique to improve the light detection and is being considered for LEGEND-1000. However, a Xe-doped LAr (\XeD) volume of sufficient size to demonstrate its use for LEGEND has not yet been demonstrated. It is important to demonstrate that the doping can be done at a large scale in a controlled manner, and to determine if \XeD\ mixture is stable.


\subsection{Xenon Doped Argon}
Doping LAr with 0.1-1000~ppm of Xe as shown by~\cite{Neumeier_2015} (10-1000ppm Xe as shown by~\cite{xenonDopedPulseShape1,Pieffer,CGWahl,Neumeier,ISHIDA1997380}) shifts the wavelength of emitted scintillation light from 128~nm, where LAr scintillates, to 175 nm where Xe scintillates. This wavelength shift is advantageous for several reasons. Argon has a long attenuation length at 175~nm~\cite{Neumeier} but only a 50-60 cm attenuation length at 128~nm~\cite{NeumeierLArAtten_2015,ISHIDA1997380}. Additionally, devices sensitive to detect both 128~nm and 175~nm are more sensitive at 175~nm.  In order to detect the 128~nm light with conventional photo-multipliers (PMT), the light must first be wavelength shifted. The coupling of light detection devices to wavelength shifting materials (e.g. TPB \cite{tpb}) typically occurs at the surface of the device, and therefore nearly half the light does not directly enter the PMT because the secondary re-emission process is isotropic. In contrast, the \XeD\ wavelength shifting occurs approximately at the interaction point. 

Obviously there are many advantages to a \XeD\ LNGD but in order to accurately quantify them, \XeD\ needs to be demonstrated on a large volume. The largest active volume previously used was 13.6~l~\cite{Pieffer}, the other active volumes used in previous works were 31~mL~\cite{xenonDopedPulseShape1}, 56~mL~\cite{NeumeierLArAtten_2015}, 1.6~l~\cite{ISHIDA1997380}, and 3.14~l~\cite{CGWahl}. One concern is the possibility of clumping: reference~\cite{ISHIDA1997380} (1.6~l active volume) claims to measure clumping at Xe concentrations of 3\%, while a chemical analysis shows that Xe is 16\% soluble in LAr at 87~K \cite{ArgonNitrogenXenon}. Another concern is the stability of the mixture, since Xe has a much higher freezing point than Ar (161.40~K versus 83.81~K).

Xenon doping was performed in previous smaller experiments using one of two methods. Either a room temperature argon-xenon mixture was injected into the liquid volume, or the mixture was prepared in the gaseous phase and then condensed into liquid. The warm gas injection technique will not work on large scale experiments because xenon homogeneity will take a long time to be achieved, while a pre-mixed gas is not feasible due to the large gas storage volumes required. Additionally, these doping techniques result in large total-concentration uncertainties making reported scintillation time constants difficult to compare. The smallest doping concentration uncertainty reported to date is 23.5\% \cite{Pieffer}. New injection techniques need to be considered if xenon doping is to be pursued on a large scale.

A large Xe doping test stand is also critical for understanding the effects of the long attenuation length at 175~nm in \XeD. This long attenuation length can result in an increase in light yield as observed previously by Ref.~\cite{Pieffer} ($\times$1.4 increase at 300$\pm$80 ppm) and Ref.~\cite{CGWahl} ($\times$1.25 increase at 1093$\pm$547 ppm). The discrepancies in these increased light yield measurements can be attributed to different detectors sizes, with the larger increase being measured in the larger detector.

%% file: 1_ArgonandXenonDopingScintillationFeatures.tex
\section{Argon and Xenon Scintillation}
\label{sec XenonDopingTheory}

When ionizing radiation interacts with a noble gas, dimers form which, when relaxing back to the ground state, produce scintillation light. Scintillation light is emitted through the transition from one of the two lowest molecular excited states, $^{3}\Sigma^{+}_{u}$ (triplet) or $^{1}\Sigma^{+}_{u}$ (singlet), to the ground state $^{1}\Sigma^{+}_{g}$. These states cannot be easily distinguished spectroscopically but do have different relaxation times. The \singlet state for liquid argon has a half life of around 5 ns while the \triplet state half lives for pure liquid argon have been measured to be between 1.2-1.6 $\mu$s (\cite{triplet1.6},\cite{triplet1.46},\cite{triplet1.26}). This large difference in measured \triplet state lifetimes for \say{pure} argon is likely due to small concentrations ($\sim$ 1 ppm) of residual atmospheric xenon\footnote{Atmospheric argon concentrations are 0.934\% by volume in Earth's atmosphere, while atmospheric xenon concentrations are 87$\pm$1 ppb \cite{nobleGasConcentrations} by volume. If the argon distillation process does not remove any xenon, as much as 9.3 ppm of xenon can remain. }. The \singlet  state for liquid xenon has a half life of 2-4 ns and the \triplet state 21-28 ns \cite{xenonEmission}. The main feature of the argon and xenon scintillation spectra is the Gaussian feature centered at 128 nm and 175 nm for argon and xenon respectively. The summed probability distribution function for the \singlet and \triplet components can be written as:
\begin{equation}
\label{eq TripletSinglet}
    I(t) = \frac{A_{1}}{\tau_1}e^{\sfrac{-t}{\tau_1}} + \frac{A_3}{\tau_3}e^{\sfrac{-t}{\tau_3}}
\end{equation}
where $\tau_{1}$ and $\tau_{3}$ are the time constants for the \singlet and \triplet  state respectively and $A_{1}$ and $A_{3}$ are the relative intensities of each components. The integral of $I(t)$ over all time is normalized to unity and therefore $A_1+A_3 = 1$.

\subsection{Xenon Doping}
\label{section XeD}

When Xe is inserted into LAr, it quenches the Ar dimers through collisional de-excitation. Unlike other contaminates (e.g. N), this collision allows for effective energy transfer between the Ar dimer and Xe, allowing the Xe atom to create a dimer of its own, as:
\begin{subequations}
    \label{equation xenonDoping}
    \begin{align}
    &\ce{Ar2^{\ast}}+\ce{Xe} + \ce{collision} \rightarrow \left(\ce{ArXe}\right)^{\ast} + \ce{Ar} \\
    & \left(\ce{ArXe}\right)^{\ast} + \ce{Xe} + \ce{collision} \rightarrow \ce{Xe2^{\ast}} +\ce{Ar}
    \end{align}
\end{subequations}
Here collision refers to a molecular collision between a Xe atom and a $\rm Ar_2^\ast$ or $\rm ArXe^\ast$ dimer. Based on estimates of the diffusion rate from \cite{fields2020kinetic}, this WLS process occurs no more than 500~$\mu$m from the initial argon interaction for Xe concentrations considered in this work.

Xe and Ar have very different scintillation time profiles with the biggest difference being their \triplet state lifetimes ($\sim$1~$\mu$s versus $\sim$30~ns). When Xe is injected into Ar at low concentrations, the \triplet state of the Ar is quenched by collisions, thus the time distribution can be described~\cite{xenonDopedPulseShape1}:
\begin{equation}
\label{eq xenonDoped scint}
    I(t) = \frac{A_{f}}{\tau_{f}}e^{-t/\tau_{f}}+\frac{A_{s}}{\tau_{s}}e^{-t/\tau_{s}} - \frac{A_{d}}{\tau_{d}}e^{-t/\tau_{d}}
\end{equation}
where $\tau_{f}$ and $\tau_{s}$ are the decay times of the fast and slow components of the Ar-Xe mixture respectively, $\tau_{d}$ is a time characterizing the conversion of $\rm Ar_2^\ast$ to $\rm Xe_2^\ast$ . $A_{f}$, $A_{s}$ and $A_{d}$ are the intensities of these three terms. In order to normalize Equation \ref{eq xenonDoped scint}, the following relation is defined: $A_{f}+A_{s}-A_{d}=1$. These time constants have been measured by several groups and the summary of these results can be found in Ref.~\cite{xenonDopedPulseShape1}. These measurements have unresolved tension, which is likely due to the Xe doping techniques and/or residual Xe or contaminants in the Ar. These groups also consider higher Xe concentrations which are not considered by this work.

\FloatBarrier

%% file: 2_experimental.tex
\section{Experimental Setup}
\label{section experiment}

\subsection{Vacuum, Cryogenics, and Recirculation Systems}
\begin{figure}
\centering
\captionsetup{width=1\linewidth}
\includegraphics[width=0.49\linewidth]{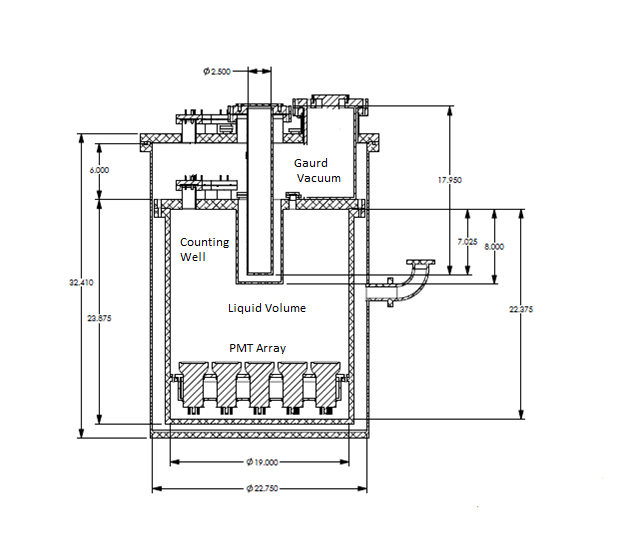}
\includegraphics[width=.49\linewidth]{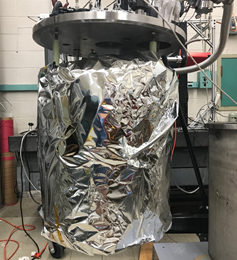}
\caption{(left) The cryostat design. 
(right) A photograph the inner vessel wrapped in aluminized infrared-reflective nylon.}
\label{fig BACoN}
\end{figure}

Our stainless steel cryostat consists of two nested cylinders (inner (IV) and outer (OV) vessels) (Fig.~\ref{fig BACoN}). The IV can hold approximately 100~l of liquid. The space between IV and OV is held near vacuum to prevent convective heating. An aluminized infrared-reflective wrapping (Fig.~\ref{fig BACoN}, right) surrounds the inner vessel to reduce radiative heating from the warm outer can. The inner vessel is attached to the lid of the outer vessel with four G-10~\cite{G10} rods.

The vessel is drip cooled using an AL60 Cryomech cold head~\cite{coldhead} providing approximately 60~W of cooling power at 80~K. The cold head is mounted on top of the OV and is connected to the IV through a thin walled, 2.75-in ConFlat bellows. The cold head drips condensed Ar down a thin stainless steel rod directly to the bottom of the inner vessel. The cold head is surrounded by a Teflon funnel, which is used to separate the warm and cold Ar gas. As Ar evaporates, it flows up the funnel toward the cold head, re-condenses, and drips back into the liquid. Warm Ar gas is inserted above the cold head and falls as it cools. The cold head is instrumented with a 50~W heater and a thermocouple, which maintains the cold head at LAr temperature.

The inner vessel was filled from the ullage of a research grade LAr cylinder\footnote{upper limit from vendor for research grade LAr states purity to be at least 99.999\% pure}. Both inner and outer vessels are leak checked prior to each fill. No thorough vessel pump and bake cycling was found necessary  as a result of our re-circulation system described below. Gas was flowed through a SAES PS4-MT3/15-R getter, which can purify noble gasses to less than 1~ppb of all common contaminates, but not noble gases~\cite{getter}. The quantity of Ar in the inner vessel is measured using a shipping scale accurate to $\pm$0.5~lbs. The mass of LAr in the inner vessel is measured as the vessel filled at a liquification rate of 0.58~kg/hr. This implies a net cooling power of about 44~W, assuming the Ar gas started at a temperature of 300~K and is liquified at 87.15~K. 

In order to purify the liquid argon after the initial fill, gaseous argon in the ullage of the IV is pumped through the getter using a stainless steel bellows pump that uses PTFE and stainless steel for all wetted surfaces. The flow is controlled using a mass flow controller \cite{flowMeter} and set to 6 slpm to match the optimal flow rate of the getter. The flow rate fluctuates no more than 1\% around the set point. The purified gas is then condensed again at the coldhead. The ullage pressure was initially set to 550 torr but then raised to 640 torr (described in Section \ref{sec NitrogenDoping}). At both pressures, the weight of the LAr and pressure of the GAr in the ullage do not change while recirculating. This implies that our coldhead is able to supply enough power to match the heat load of recirculation system. 

\subsection{Gas Doping System}

\begin{figure}[ht]
\centering
\captionsetup{width=1\linewidth}
\includegraphics[width=0.8\linewidth]{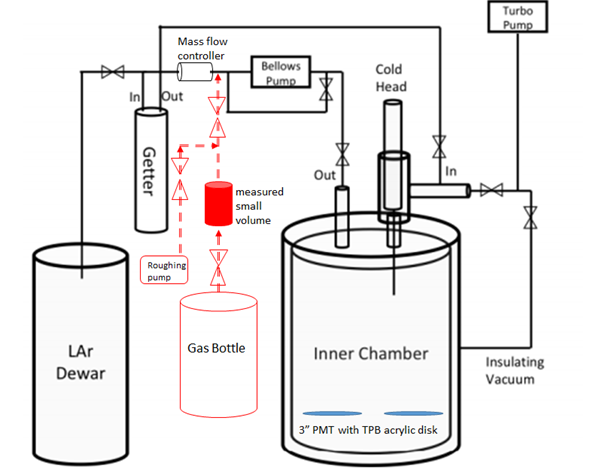}
\caption{Piping diagram for the gas insertions.} 
\label{fig BACoNPiping}
\end{figure}

The gas doping system consists of a small measured cylindrical volume, 19$\pm$1~ml, connected to a supply gas bottle on one side and the recirculation system on the other (Fig.~\ref{fig BACoNPiping}). The gas pressure is measured using a high pressure Baratron gauge with a precision of 1~Torr and a range from 0 to 2000~Torr. Prior to any doping run, the doping volume is pumped out using a turbo pump to ensure cleanliness. The doping gas is inserted in steps of $\sim$0.5~ppm ($\sim$1500~Torr) with the largest achievable step size of $\sim$0.9~ppm. Larger doping steps are not considered due to the constraints from the maximum range of the high pressure gauge and the small size of the doping volume. The pressure in the doping volume is chosen to be larger than that of the recirculation system to ensure the gases mix adequately once the doping volume is exposed to the recirculation system. The remaining gas in the doping volume is expected to mix based on the diffusion rate of the binary gas mixture (\cite{HWATT,Zhijing}), exposure time and small doping volume. About 30 minutes pass between each doping step to ensure that the doping volume has had time to fully mix with the recirculated gas. The doping volume is evacuated before each doping step using a dry scroll pump down to a pressure of 10~mTorr. We estimate that a negligible amount of gas is lost while evacuating the doping volume. During the last doping step, the doping volume is over pressured and the dopant is slowly leaked into the vacuum system using a \sfrac{1}{4}-in bellows valve until the desired pressure is reached. This last step gives excellent control over the total amount of dopant inserted per run.

It is assumed that given enough time, the dopant inserted in the gas will adequately mix with the LAr through diffusion at the liquid/gas interface. This hypothesis is validated from the results N doping (see Section \ref{sec NitrogenDoping}). It is known to the aurthors that dopants will have different equilibrium concentrations in GAr versus LAr. For example, it is known that the vapor pressure of N in GAr is about three times larger than in LAr \cite{triplet1.26} while vapor pressure of Xe in GAr is about 4000 times less than in LAr \cite{fields2020kinetic}. In either case, as long as the ratio of the dopant's vapor pressure in GAr compared to LAr is much smaller than the ratio of the GAr mass to LAr mass ($\sim$4000 for our case), it is assumed that almost all of the dopant will dissolve in the liquid. For these measurements, only dopants that are soluble in LAr are chosen.

The use of high purity gases ensures that the number of mols of the dopant could be calculated using the ideal gas law (PV = nRT). Molar concentration in ppm can then be calculated by using the total weight of the liquid argon volume measured using a shipping scale (119$\pm$0.2 kg). The measured temperature of the lab room ($\sim$~295~K), after the doping system has reached equilibrium, is taken as the doping gas temperature. This temperature typically fluctuates about 2 K around the set point of the room thermostat. As mentioned above, the measured pressure fluctuates by about 1~Torr for each doping step.  Each doping step has approximately the same measurement uncertainty ($\sim$5\%) due to insertion of gas at the same pressure, volume, and temperature. Hence, as the total number of doping cycles increased, the relative measurement error on the total concentration decreases, causing the total error to be dominated by the systematic error of the doping volume. 

\subsection{Electronic Systems}

The inner vessel is instrumented with two 12-stage Hamamatsu, 3-in PMTs (R11065)~\cite{pmtR11065} which are box\&linear-focused, with synthetic silica 3-in windows and bi-alkali photo-cathodes designed to operate at LAr temperatures. The PMT cathode ground scheme is chosen to allow signals and high voltage to share a cable, reducing the heat load from the wires. Each PMT is instrumented with a fiber optic cable and a blue LED which are used to check the stability of the PMTs. One PMT was lost while the cryostat was being filled with LAr though the other PMT remained stable and had constant gain through out data taking. The PMT signals are amplified using a Phillips Scientific Octal Variable gain amplifier Nim Model 777~\cite{phillipsScientific}. The gain of our single PMT was estimated using dark count to be $(4.64\pm1.45)\times10^{6}$ after accounting for the $\times$2.86 amplification of the pre-amp circuit. The conflat feed-throughs are mounted on the warm side of the cryostat to ensure thermal cycling did not damage the ceramic/steel interface. Signals are digitized using a Tetronix TDS 3052B digital oscilloscope and read out to a laptop using custom Python software. The scope has a slow read-out rate of $\sim$1~Hz though it can digitize at 1~GS/s allowing for 1~ns bins over 10~$\mu$s.

\subsection{Pulse Finding}
\begin{figure}
\centering
\captionsetup{width=1\linewidth}
\includegraphics[width=1.0\linewidth]{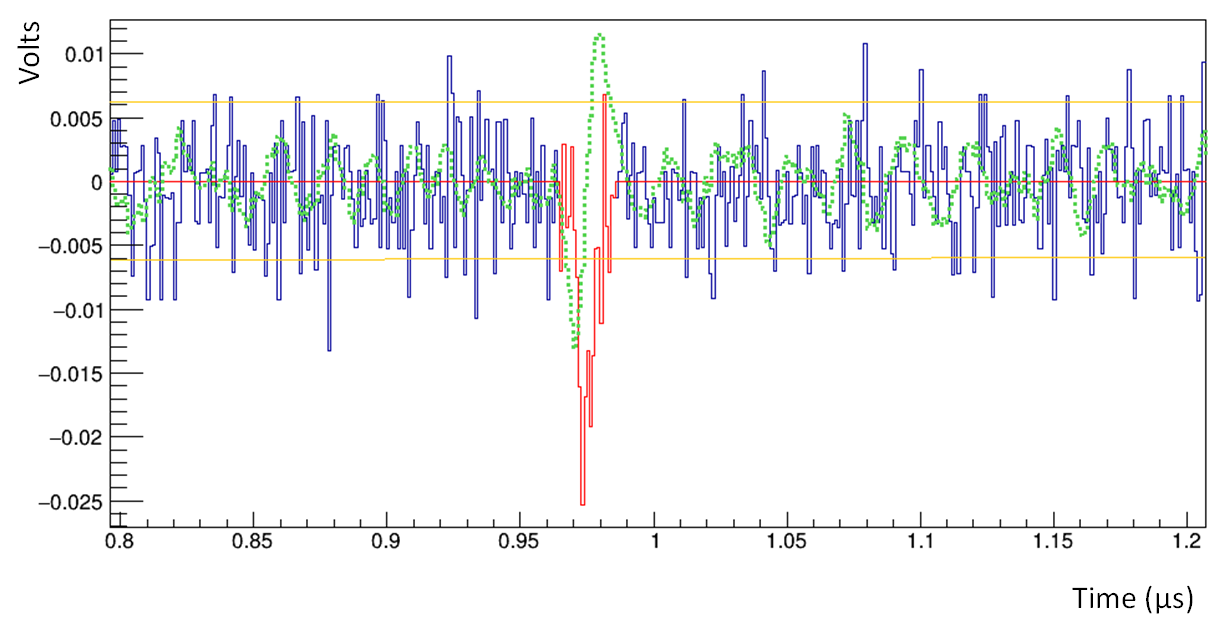}
\caption[Pulse Finding Example]{
A single photon pulse found using the pulse finding algorithm~\cite{Zugec:2016}. The solid blue histogram is the background noise while open red histogram is the highlighted found pulse. The thick dashed green line is the derivative of the blue plus red lines (signal plus background). The two yellow lines are the 3.5 $\sigma_{rms}$ derivative thresholds for the  crossings, with two crossings per threshold required for a pulse.} 
\label{fig PulseFindExample}
\end{figure}

A pulse finding method was adapted from Ref.~\cite{Zugec:2016}, which uses a smoothed derivative method to identify signal pulses as large (3.5$\sigma$) fluctuations in the noise. Four threshold crossings of the derivative define a pulse.  An example of this pulse finding can be seen in Fig.~\ref{fig PulseFindExample}. Local minima are identified using the derivative allowing individual pulses to be distinguished in the event of pulse pileup. This feature proved essential for our measurement of 
the \triplet lifetime of Ar. 

\subsection{TPB Coating}

The PMT windows have a mounted tetraphenyl butadiene (TPB) coated acrylic disk to wavelength shift the Ar scintillation light into the sensitive region of the PMTs. To prevent the PMTs from floating in the LAr, the TPB disks are tied down to the Al holders with stainless steel wire. The average quantum efficiency over the TPB emission spectrum is $\sim$29\%, making it well suited for measuring VUV scintillation \cite{Acciarri_2012}. TPB is vapor deposited to a thickness of about 2~$\mu$m. The QE of the TPB coating was never measured because we planned to do a relative light yield measurement. 

\begin{figure}
\centering
\captionsetup{width=1\linewidth}
\includegraphics[width=.3675\linewidth,height=.49\linewidth,angle=90]{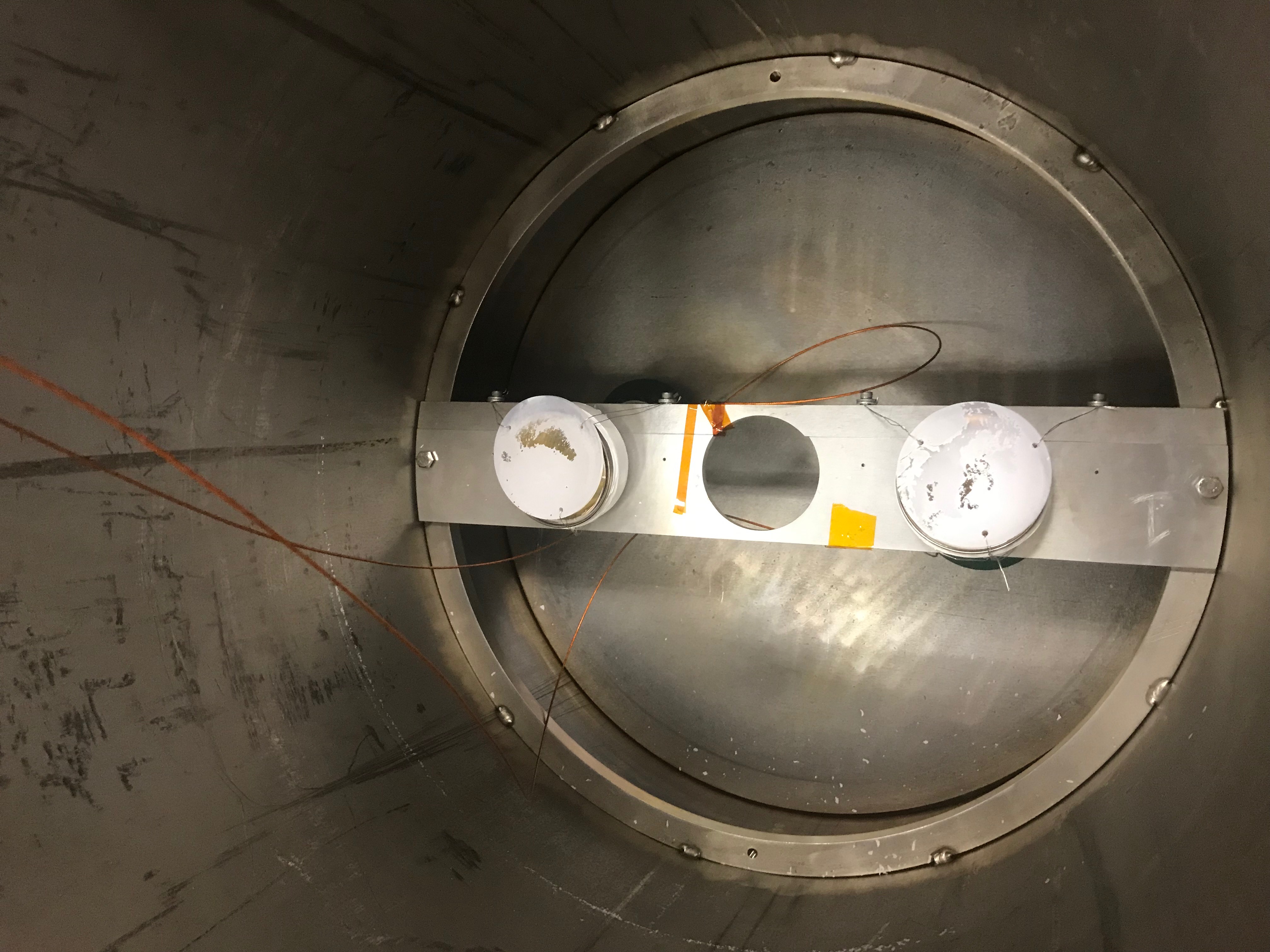}
\caption[]{PMTs mounted with TPB coated acrylic disks inside IV after an initial test fill. A ring of TPB can be seen on the left side of the IV above the welded steel ring. Small TPB particulate can be seen on the bottom of the IV.} 
\label{fig TPBFailure}
\end{figure}

After completing an initial test run, it was found that TPB from the acrylic disks had dislodged and redeposited in a ring on the wall of the cryostat at the approximate height of the PMT windows (Figure \ref{fig TPBFailure}). Additionally there were pieces scattered on the bottom of the IV that were likely floating on top of the LAr. Though TPB has been shown to be stable in liquid argon~\cite{Gerda}, it is speculated that the volatile boundary between the gas and liquid phases of the Ar was abrasive to the vapor deposited TPB. As the liquid level rose above the PMTs, no further abrasion was observed. If this abrasion happened in for the data presented in this paper, it is not known what effect this has on the light yield of the TPB. Light from the ring of TPB should not significantly contribute to the light seen in the PMT because the TPB ring is at the horizon of the PMT's view. WLS light from the suspected floating TPB particulate likely could be seen. We estimate that this WLS light contributes negligibly to the total light seen because of the small relative area of the TPB flakes when compared to the total surface area of the liquid. 

%
%

%

%% file: 3_Triplet-Lifetime.tex
\section{Triplet Lifetime Measurement}
\label{sec tripletLifetime}

\subsection{After-pulsing}
\begin{figure}[ht]
\centering
\captionsetup{width=1\linewidth}
\includegraphics[width=1.0\linewidth]{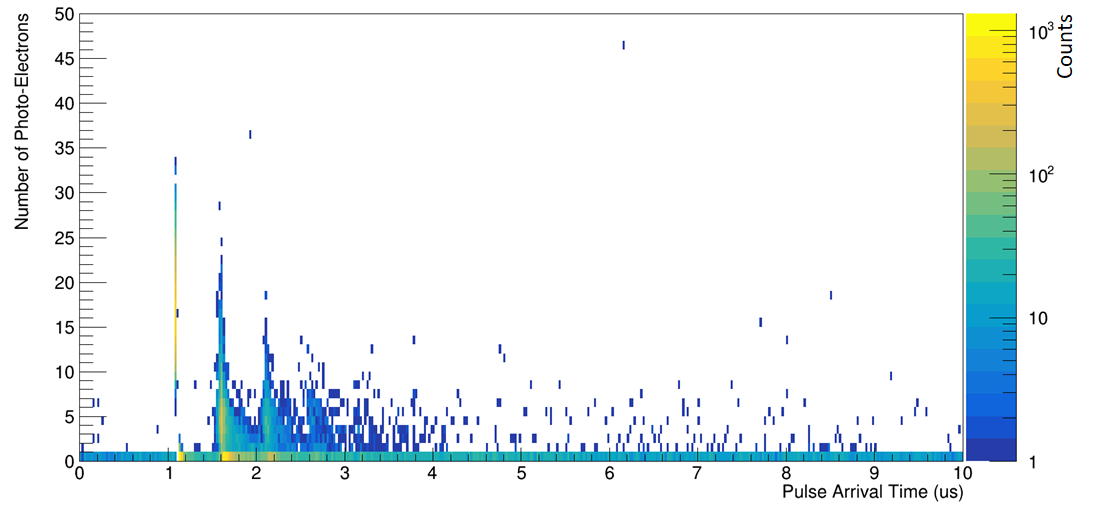}
\caption[]{A large LED pulse can be seen at $\sim$1~$\mu$~s and after pulses can be seen peaking at 1.6 $\mu$s, 2.1 $\mu$s, and 2.6 $\mu$s }
\label{fig After-Pulse}
\end{figure}

After-pulsing was noticed in the data between 1 $\mu$s and 3 $\mu$s. In order to understand the effects of after-pulsing, a large LED pulse was used to generate after-pulses. The charge weighted pulse arrival time distribution of the after pulses can be seen in Figure \ref{fig After-Pulse}. The first after-pulse has an area that is 20.5\% of the initial peak, the second pulse has an area that is 5.2\% of the initial peak and the third after-pulse has an area of about 1.5\% of the initial peak. The exact area of third after-pulse is harder to determine because the dark count has a large contribution to its total area. The effects of after-pulsing cannot be reduced without decreasing the single photon resolution and thus the nominal operation voltage of 1600~V was chosen. 

Based on the literature describing after-pulsing (\cite{afterpulse:2017twm}), it is clear that after-pulsing will most likely be triggered from the singlet component of the scintillation light due to the large size and short duration of the initial signal. In contrast, the probability of triggering an after-pulse from a single photo electron is estimated to be small ($<1\%$) based on the number of large events seen in dark rate calibration data. The relative contribution of after-pulses should not change with dopant concentration because the singlet intensity does not change with at these concentrations. Thus when doing a light yield comparison, the effects of after-pulsing will not drop out when looking at the relative ratio of photo-electrons before and after doping but rather be a constant effect regardless of doping concentration.

\begin{figure}[ht]
\centering
\captionsetup{width=1\linewidth}
\includegraphics[width=1.0\linewidth]{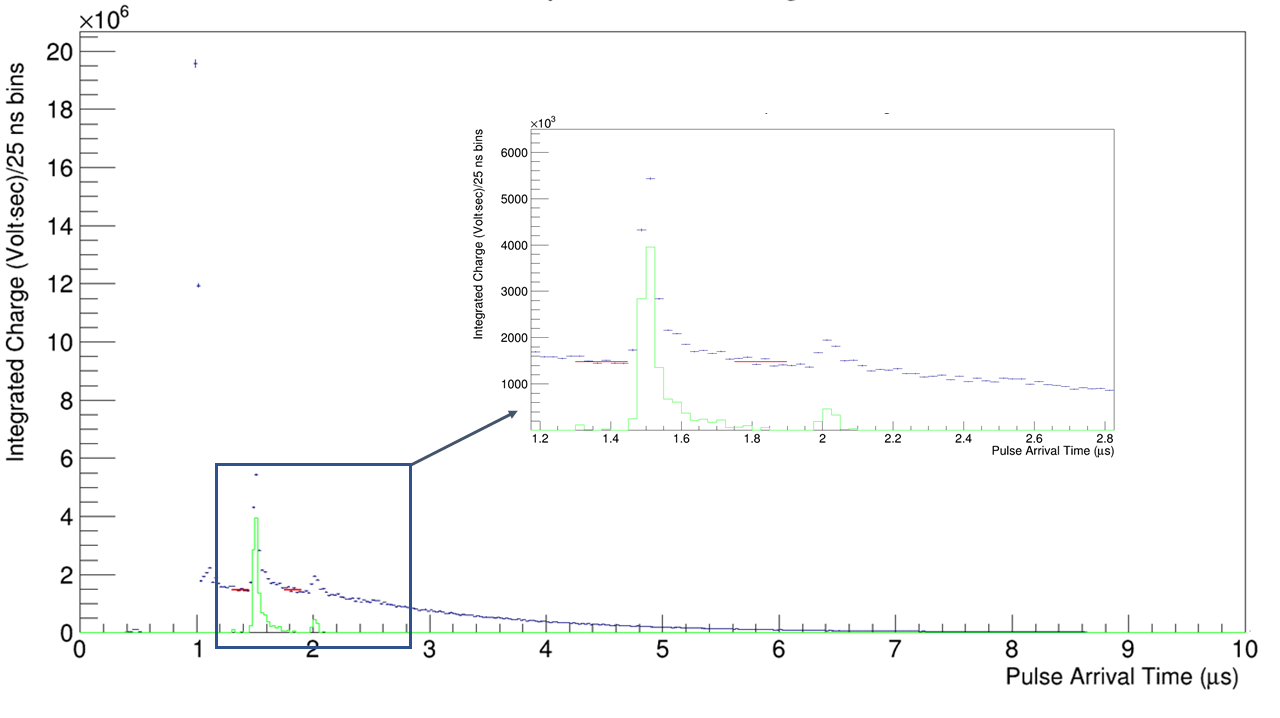}
\caption[]{An example of how after-pulsing is estimated using an exponential fit. The blue markers are data, while the solid red line is the exponential fit to the side bands of the first after-pulse peak, and the solid green histogram is the waveform subtracted estimate of the after-pulses. A zoom in of the after pulse region is shown in the insert.}
\label{fig After-Pulse-estimate}
\end{figure}

To estimate the contribution of after-pulsing to the total charge of a distribution, an exponential is fit in two side-bands windows around the first after-pulse peak. The windows are chosen to be 1.30 to 1.45 $\mu$s and 1.75 to 1.90 $\mu$s. This exponential is assumed to be the underlying distribution around the first after-pulse peak. The fit is then subtracted from the waveform leaving behind an estimate of the after-pulse. An example of this procedure performed on pure LAr data can be seen in Figure \ref{fig After-Pulse-estimate}. Through this technique, the contribution of after-pulses to the total charge of a waveform can be made and later added in as a systemic effect when reporting on light yield. Finally, it is worth mentioning that an attempt to subtract out the after-pulsing peaks using the shapes found in the after-pulsing calibration data was performed. This method ultimately failed because although the shapes appear similar by eye, they do not match well enough to adequately remove the after-pulsing without distorting the distributions.

\subsection{Fitting the Triplet Lifetime} 

We investigated mitigating the after-pulsing effect on the LAr \triplet lifetime using two fitting methods. In both methods the singlet was not fit because the time resolution of the initial trigger was too large. Data was taken using a random trigger in which the trigger level was set to $\sim15$~PE.

\begin{figure}[]
\centering
\captionsetup{width=1\linewidth}
\includegraphics[width=.49\linewidth]{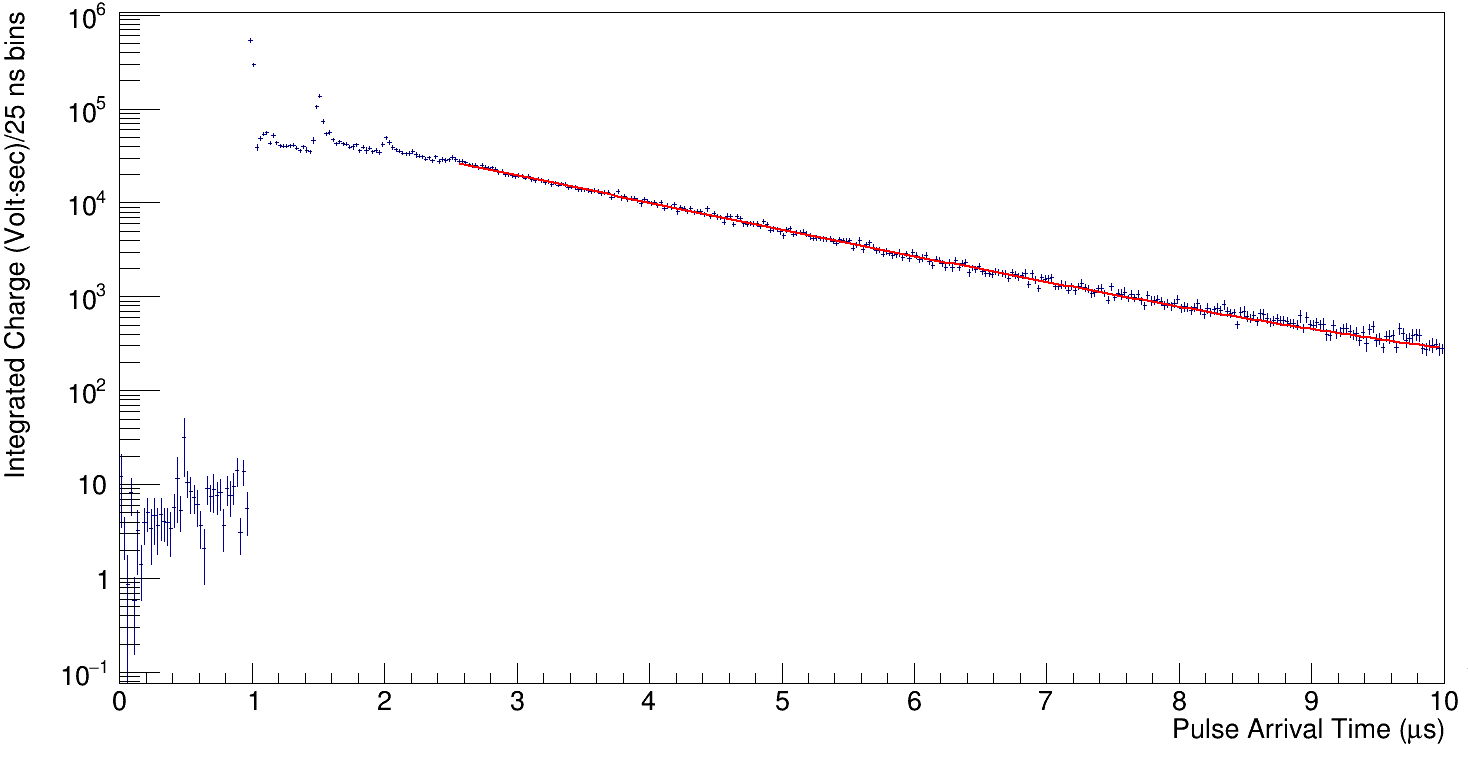}
\includegraphics[width=.49\linewidth]{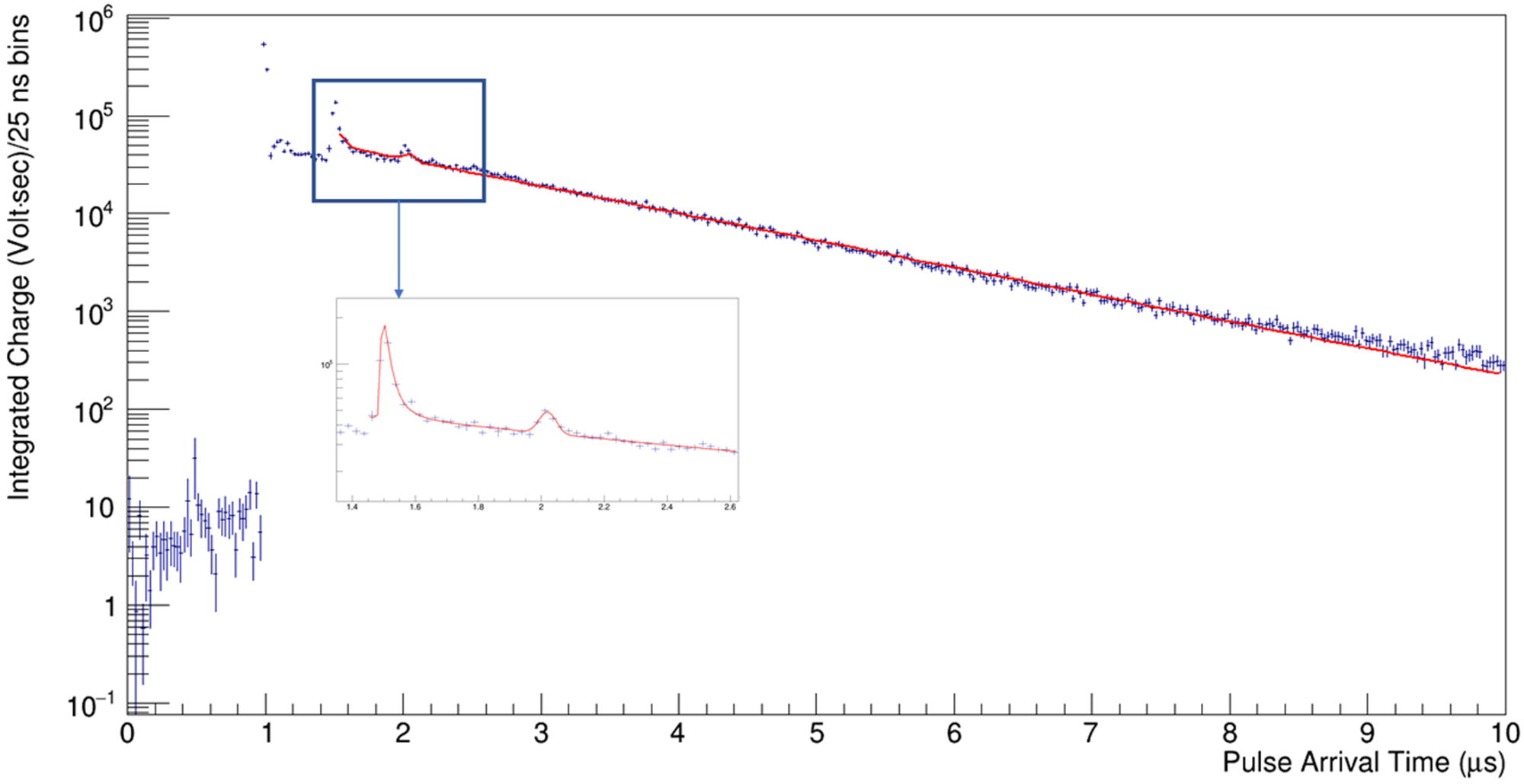}
\caption[]{Two fits to the \triplet scintillation distribution in pure LAr. The left plot is the fit with a single exponential and a constant from 2.5~$\mu$s to 10~$\mu$s. The right plot shows the fit with a Landau shape for the first after-pulse and a Gaussian shape for the second after-pulse with an exponential and a constant fit the tail. The fit to the after-pulsing region is shown in the insert. } 
\label{fig afterPulseFitting}
\end{figure}

In the first fitting method, the \triplet lifetime was found using a single exponential plus a constant fit. A constant is added to account for GAr scintillation light coming from the ullage. A start time of 2.5~$\mu$s was chosen in order to avoid the first two the after-pulses. A \triplet lifetime of $1456\pm38$~ns was measured with this method. Alternatively, we fit a Landau shape to the first after-pulse and a Gaussian shape to the second after-pulse. These functions are added to the first function and the sum was fit starting from 1.75 $\mu$s. The average $\sfrac{\chi^2}{NDF}$ for this second fitting method was $1.44\pm0.12$ indicating that this chose of functions is reasonable. This method yields a systematically larger value of the \triplet lifetime of $\rm 1502\pm30 ns$. An example of these two different fitting methods can be seen in Fig.~\ref{fig afterPulseFitting}. It is likely that this second fitting method yields a larger value because the two after pulse function account for some of the light at early times thus flattening the exponential component of the fit. An average was taken as the true \triplet lifetime, measured to be $\rm 1479\pm 38 (stat) \pm 23 (sys) \; ns$ where the statistical uncertainty was taken to be the larger of the two different fitting method results and the systematic uncertainty was taken to be difference in the two fitted \triplet lifetimes. It is estimated, using the technique mentioned above, that after-pulsing contributes about 6.4\% to the total charge.

\FloatBarrier

%% file: 4_NitrogenDoping.tex
\section{Nitrogen Doping}
\label{sec NitrogenDoping}

\subsection{Measuring Birk's Constant}
In order to validate the gas doping and recirculation system, a small measured quantity of N$_2$ was injected using the methods described in Section \ref{section experiment}. Following the injection, the N$_2$ was filtered out using the getter. The effects of N on Ar scintillation light are well known and have been characterized in Refs.~\cite{triplet1.26} and \cite{Himi:1982hf}. The relationship between the measured triplet lifetime ($\tau_{3}'$) and the true triplet life time ($\tau_{3}$) follows Birk's law:
\begin{equation}
\label{birks law}
    \tau_{3} ' = \frac{\tau_{3} }{1 + \tau_{3} \cdot k \cdot C}
\end{equation}
where $k$ is Birk's constant and $C$ is the molar concentration of impurities measured in parts per million. Birk's constant has been measured to be 0.11 ppm$^{-1} \mu$s$^{-1}$ and $0.13$ ppm$^{-1} \mu $s$^{-1}$ for N in LAr by the two previously mentioned authors. They obtained  $\tau_{3}$ = 1.26 $\mu s$ compared to our value of  1.48 $\mu s$. 
Reference~\cite{triplet1.26} was aware that their $\tau_{3}$ was smaller than that typically measured~(\cite{triplet1.46}, 1.46 $\mu$s) and speculated that if they used the fitting method of Ref.~\cite{triplet1.46} and extrapolated their $\tau_{3}$ using Eqn.~\ref{birks law} to pure LAr, their value becomes $\rm \tau_{3}= 1.45 ~\mu s$, which is in good agreement with our measurement once errors are included.

Before doping, the recirculation and purification system was run and $\tau_{3}$  measured frequently. Once the $\tau_{3}$ measurement was stable, it was assumed that the Ar was pure.  The nitrogen contaminant was then inserted and $\tau_{3}$ decreased according Eqn.~\ref{birks law}. As the Ar was purified, it is assumed that the contamination concentration decreased as:
\begin{equation}
    C(t) = C_0 \cdot e^{\sfrac{-t}{\tau_{filter}}}
\end{equation}
where $C_0$ is the initial concentration measured in molar parts per million and $\tau_{filter}$ is a fitted time constant characterizing the filtration process.
 
\begin{figure}
\centering
\captionsetup{width=1\linewidth}
\includegraphics[width=1.\linewidth]{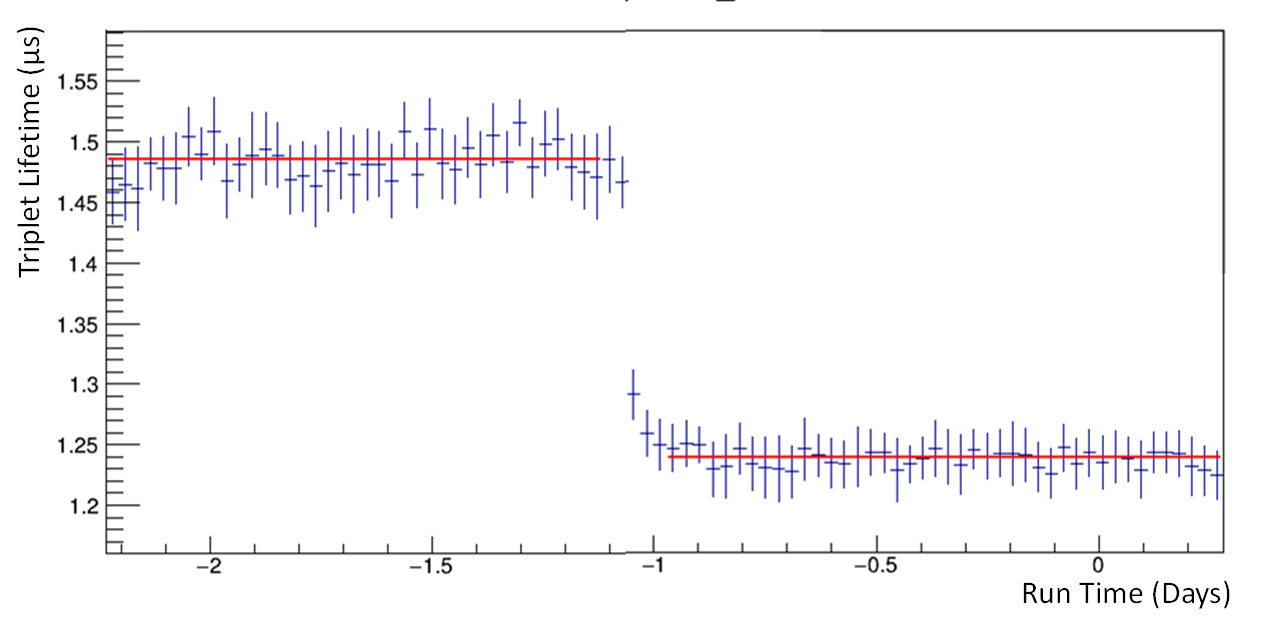}
\caption{ Triplet lifetime ($\mu$s) versus run time (days) for the nitrogen doping run. The nitrogen was injected at run time -1.1 days. It can be seen that the nitrogen thoroughly mixed after about three hours.} 
\label{fig BirksDrop}
\end{figure}

\begin{figure}
\centering
\captionsetup{width=1\linewidth}
\includegraphics[width=0.99\linewidth]{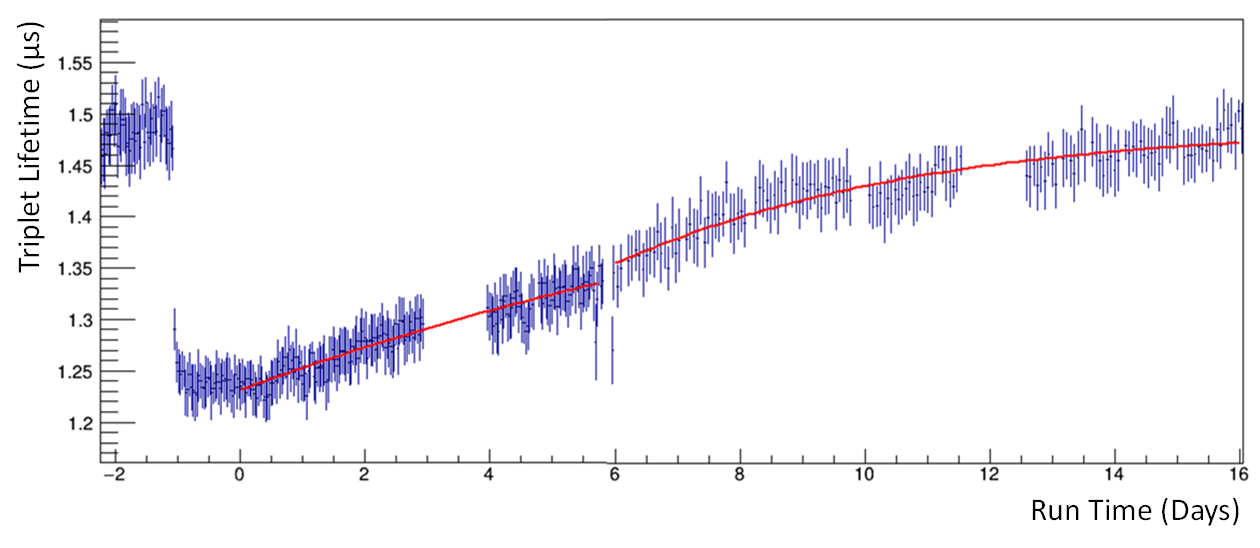}
\caption[ Nitrogen doping in pure LAr]{ Nitrogen doping in pure LAr. The left fitted red line shows the fit obtained when the heat leak for the system was minimized, yielding a filtration rate of 9.5$\pm$0.6~d. The right fitted red line shows the fit obtained when a heat leak was added to the system, yielding a filtration rate of 4.5$\pm$0.3~d.} 
\label{fig BirksFit}
\end{figure}

We performed one $N_2$ doping run injecting 1.06$\pm$0.07~ppm of N$_2$. The nitrogen was injected in two steps using the techniques outlined in Section \ref{section experiment}. The triplet lifetime dropped from $1479 \pm 38 (stat) \pm 23 (sys) \; ns$ to $1238 \pm 15 (stat) \pm 20 (sys) \; ns $ as seen in Fig.~\ref{fig BirksDrop}. The quenched triplet lifetime and associated errors are calculated in the same way as in Section \ref{sec tripletLifetime}. It is found that after-pulsing contributes about 7.7\% to the total charge seen in these quenched waveforms. From this reduced lifetime a Birk's constant of 0.12$\pm$0.03~$\rm ppm^{-1}\mu s^{-1}$ was found, consistent with the previous measurements. Using this measured value of Birk's constant, Eqn.~\ref{birks law} was fit to the triplet purification curve in order to evaluate the effectiveness of the recirculation system. Initial measurements of the filtration time constant found it to be 9.5$\pm$0.6~d (left fitted curve in Fig.~\ref{fig BirksFit}). To increase the filtration rate, the turbo pump was valved off from the guard vacuum while leaving the roughing pump on thereby decreasing the insulating vacuum from $5.0\times10^{-7}$~Torr to $1.4\times 10^{-3}$~Torr. Additionally the cold-head set point was raised from -190~C to -187~C thus increasing the internal pressure of the cryostat from 550~Torr to 640~Torr. A modification to Eqn.~\ref{birks law} was made to account the for non-zero start time to obtain:
 \begin{equation}
\label{birks lawMod}
    \tau_{3} ' = \frac{\tau_{3} ' }{1 + \tau_{3} ' \cdot k \cdot C_{0} \cdot e^{\sfrac{-(t-t_{0})}{\tau_{filter}}}}
\end{equation}
where $t_{0}$ was the non zero start time of the fit. The filtration time constant measured by fitting Eqn.~\ref{birks lawMod} to the triplet purification curve after these changes was measured to be 4.5$\pm$0.3~d (right fitted curve in Fig.~\ref{fig BirksFit}).


By monitoring the cosmic ray muon peak before and after doping (see Section~\ref{sec XeD}), it was found that the most probable value (MPV) shifted from 334$\pm$21(syst)$\pm$7(stat)~PE to 283$\pm$22(syst)$\pm$7(stat)~PE, which corresponds to a 15.3$\pm$1.5\% drop. This drop is compatible with  Ref.~\cite{triplet1.26} who measured the total quenching of LAr waveforms doped with 1 ppm N$_2$ using gammas from a $^{60}$Co source. 

\section{Xenon Doping}
\label{sec XeD}
\begin{figure}
\centering
\captionsetup{width=1\linewidth}
\includegraphics[width=1.0\linewidth]{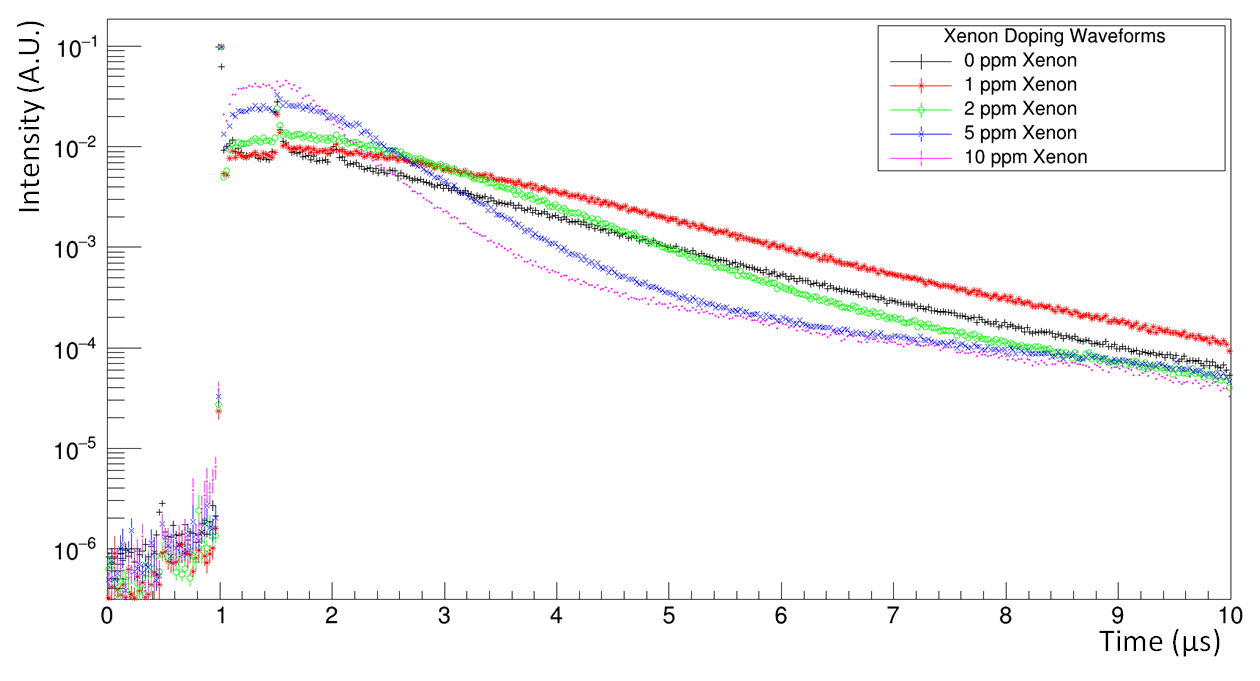}
\caption[Pulse arrival time weighted by pulse charge for xenon concentrations 1 ppm, 2 ppm, 5 ppm, and 10 ppm]{Scintillation distributions created by binning pulse arrival times weighted by the pulse charge for various concentrations of Xe from 0--10~ppm. The distributions are normalized by their integral and then the whole distribution is scaled such that their singlet contributions are the same for all doping concentrations. } 
\label{fig AveragedWaveforms}
\end{figure}

Having validated our doping procedure with nitrogen, we then used xenon. Xe doping was done in four steps of  1.00$\pm$0.06~ppm, 2.0$\pm$0.1~ppm, 5.0$\pm$0.3~ppm, and 10.0$\pm$0.5~ppm. Xenon was injected in 0.5~ppm increments until the desired molar concentration was reached. High concentrations of Xe ($\sim$100~ppm) were not considered\footnote{The largest possible insertion step is $\sim0.9$~ppm and each 0.5~ppm injection took about 30~min. Additionally, the next Xe concentration of interest is at $\sim$100 ppm where previous studies have been performed. Thus achieving this next concentration would take in total 2 days of continues doping.}. Once the concentration was stable and fully mixed, 50,000 random background events were collected at each concentration to create the scintillation distributions seen in Fig.~\ref{fig AveragedWaveforms}. Scintillation distributions are created by binning pulse arrival times weighted by the pulse charge. The waveforms were fit from 1.05~$\mu$s to 10~$\mu$s to use as much of the waveform as possible while avoiding the singlet contribution. The start time of the fit was varied by $\pm$25~ns in order to estimate the systematic uncertainty associated with the fit window. After-pulsing contributions to the total charge for 1, 2, 5, and 10 ppm were estimated to be 2.7\%, 2.5\%, 2.0\%, and $>1\%$ respectively. 

A modified version of Eqn.~\ref{eq xenonDoped scint} was used to measure the time constants of the mixture in which the fast component was removed:
\begin{equation}
\label{eq xenonFit}
    I(t) = \frac{A_s}{\tau_s} e^{-t/\tau_{s}} - \frac{A_d}{\tau_d} e^{-t/\tau_{d}} + \frac{A_{long}}{\tau_{long}} e^{-t/\tau_{long}} 
\end{equation}
where $\tau_{s}$ is the slow component, $\tau_{d}$ is the energy transfer time between the Xe and triplet state of the Ar, $A_{1}$, $A_{2}$ and $A_{3}$ are the intensities, and $\tau_{long}$ is an additional time constant to account for the exponential tail at late times. This artifact is consistent with light coming from the cold gas in the ullage \cite{fields2020kinetic}.

\begin{figure}
\begin{tabular}{cc}
  \includegraphics[width=.4950\linewidth]{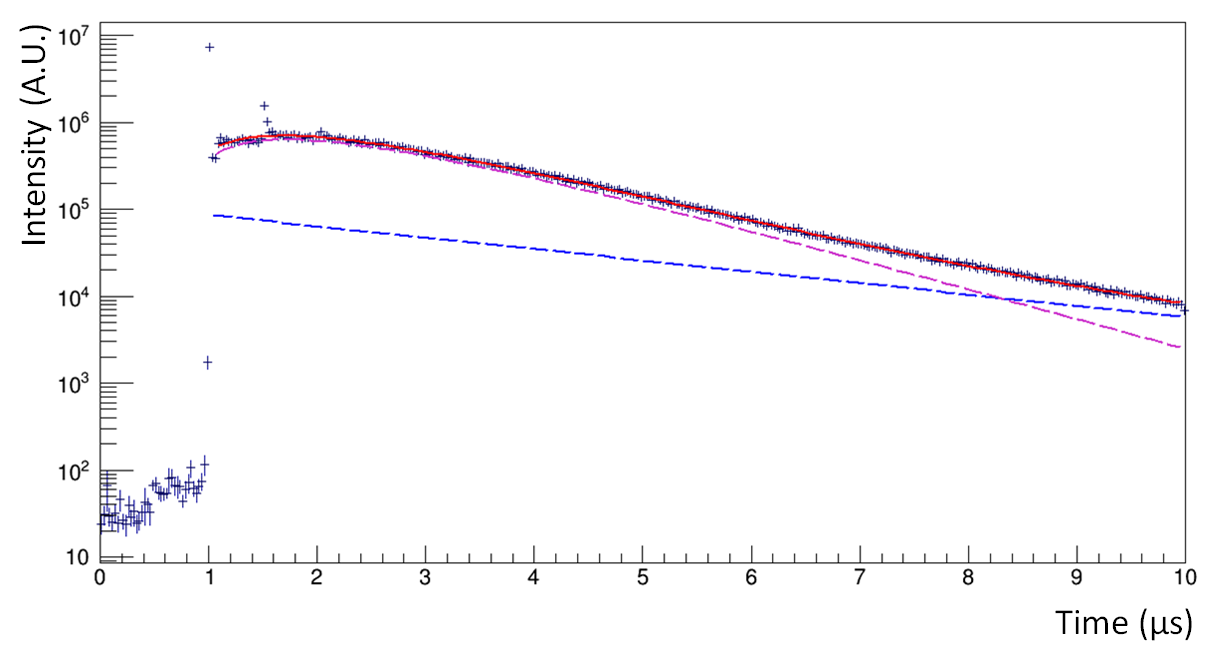} &
  \includegraphics[width=.4950\linewidth]{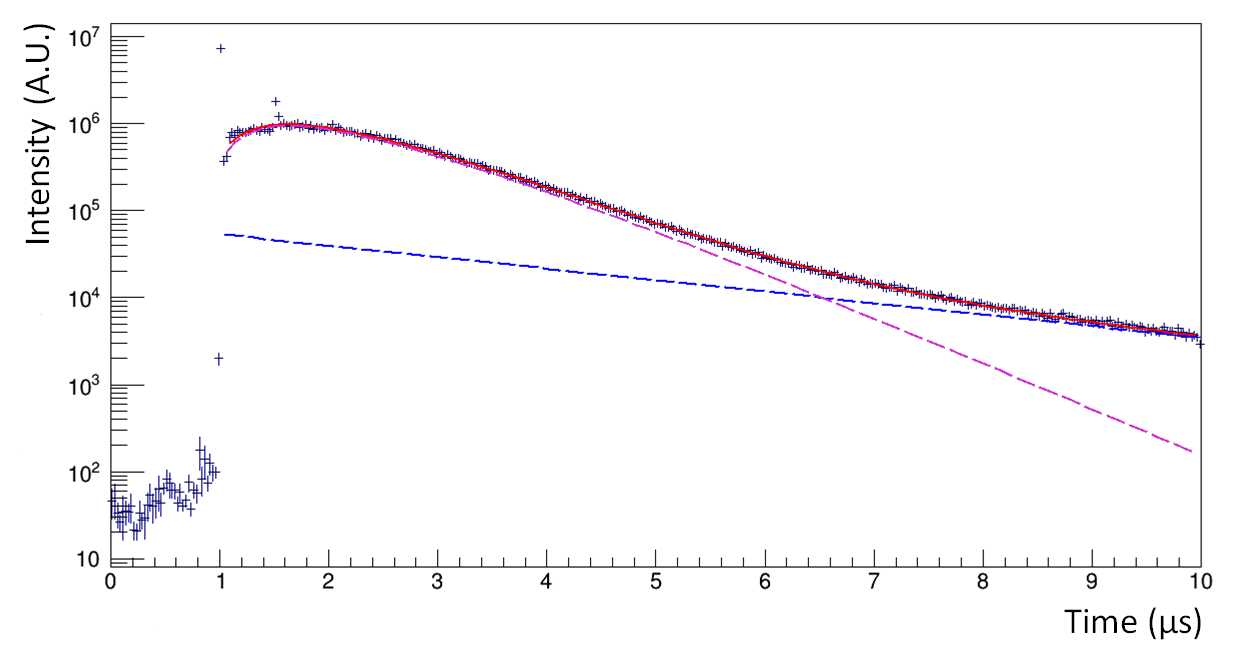}\\
(a) 1 ppm  & (b) 2 ppm \\[6pt]
 \includegraphics[width=.4950\linewidth]{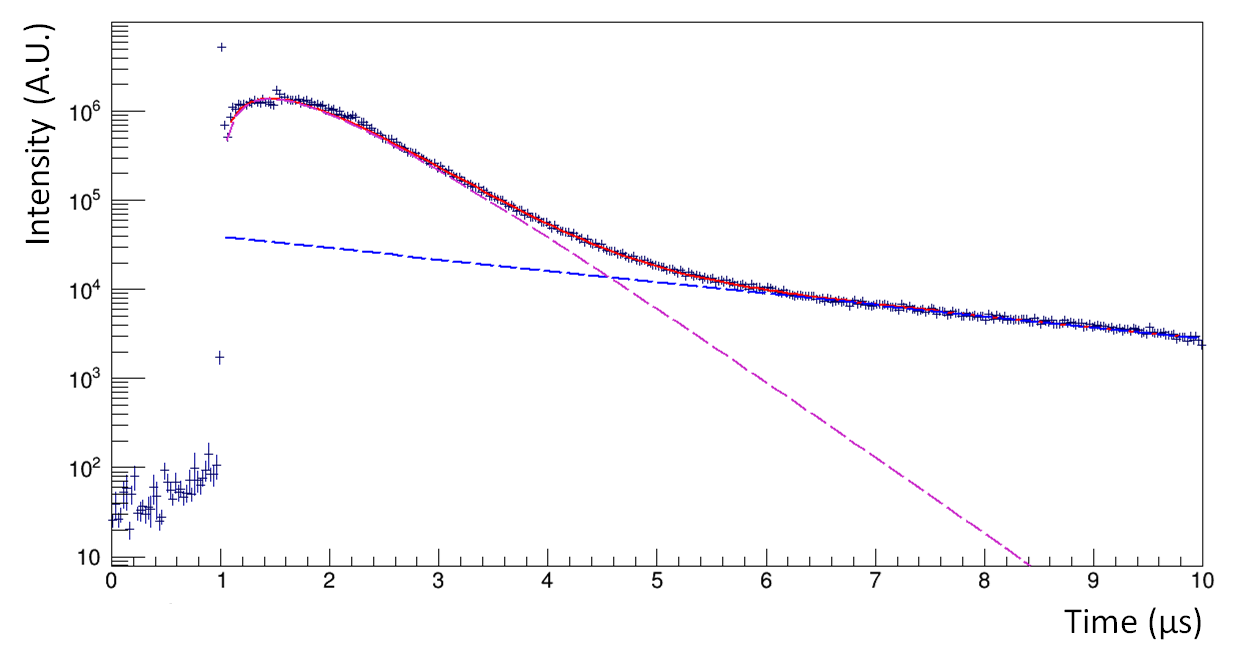} &
 \includegraphics[width=.4950\linewidth]{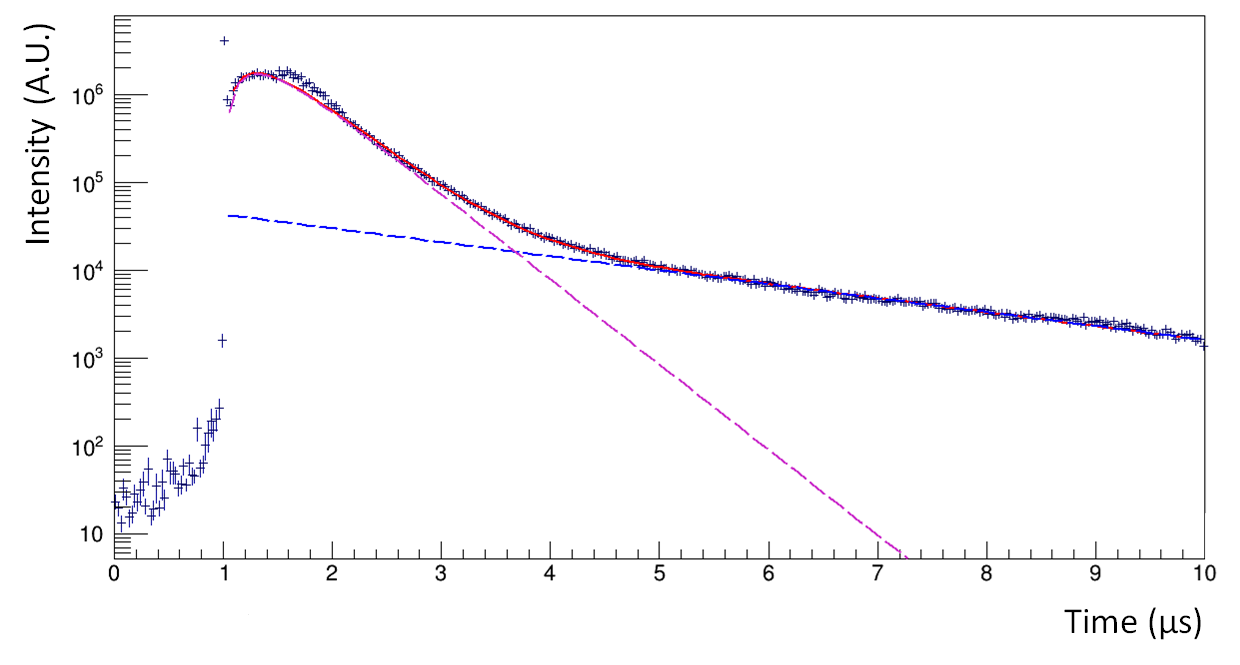}\\
(c) 5 ppm & (d) 10 ppm \\[6pt]
\end{tabular}
\caption[Fits for Xenon doping waveforms]{ Fit of Eqn.~\ref{eq xenonFit} for Xe doped waveforms with Xe concentrations of 1, 2, 5, and 10~ppm. The solid red line is the total fit, the dashed blue line is the long exponential component associated with $\tau_l$, and the magenta dashed line is the Xe dimer and Ar triplet component ($A_1 e^{-t/\tau_{s}} - A_2 e^{-t/\tau_{d}}$). Deviations from the fit at short times (1.5-2.5~$\mu$s) are due after-pulsing.}
\label{fig XenonFits}
\end{figure}

The slow component, $\tau_s$, represents the time constant for the combination of unquenched Ar dimer light and Xe dimer light created through the mechanism outline in Section~\ref{section XeD}. Though Xe decays rather quickly compared to Ar, the population of Xe dimers is {\em pumped} by the Ar-Xe mixed state, thus is able to have a much longer decay time compared to pure liquid Xe. The total concentration of Xe dimers and Ar dimers in the triplet state is $A_1 e^{-t/\tau_{s}} - A_2 e^{-t/\tau_{d}}$. The fitted waveforms are seen in Fig.~\ref{fig XenonFits} and the results from these fits are summarized in Table~\ref{table XeDTimes}. As the Xe concentration increases, the energy transfer process becomes more effective. This is reflected in the fact that $\tau_{d}$ is anti-correlated with Xe concentration. Additionally, $\tau_s$ decreases because the relative population of Xe dimers to Ar dimers in the triplet state increases with increased Xe concentration.  

\begin{table}[]
\centering
\begin{tabular}{ |p{2.75cm}||p{2.5cm}|p{2.5cm}|p{2.5cm}|  }
 \hline
 \multicolumn{4}{|c|}{Xenon Doping Time Constants} \\
 \hline
  Concentration & $\tau_{s}$ (ns)& $\tau_{d}$ (ns)& $\tau_{long}$ (ns)\\
 \hline
 1 ppm  & 1243$\pm$7$\pm$6    & 871$\pm$10$\pm$4   & 3532$\pm$212$\pm$117  \\
 2 ppm  & 771$\pm$12$\pm$1    & 721$\pm$11$\pm$1   & 3285$\pm$4$\pm$5  \\
 5 ppm  & 503$\pm$11$\pm$1    & 435$\pm$9$\pm$1    & 3415$\pm$3$\pm$19   \\
 10 ppm & 447$\pm$1$\pm$2     & 213$\pm$4$\pm$2    & 2732$\pm$4$\pm$14  \\
 \hline
\end{tabular}
\caption[Summary of Xe doping time constants]{Summary of Xe doping time constants, measured in ns, for various molar concentrations from 1 to 10~ppm~Xe. The first uncertainty is a systematic uncertainty estimated by varying the fit start time from 1.025~$\mu$s to 1.075~$\mu$s and the second uncertainty is statistical. }
\label{table XeDTimes}
\end{table}

Previous measurements~\cite{xenonDopedPulseShape1,CGWahl} of $\tau_{s}$ for 1~ppm~Xe reported values between 1~$\mu$s and 2.5~$\mu$s, though the uncertainty on the measured concentration is also 1~ppm. At 10~ppm~Xe, $\tau_{s}$ has been measured as 280~ns and 750~ns, with the concentration uncertainty at least 3~ppm. $\tau_{d}$ has not been measured for 1~ppm~Xe, but for 10~ppm the value ranges from about 250~ns to about 700~ns. Past measurements of these time constants have a much larger Xe-concentration uncertainty making them difficult to compare with our measurements. Disagreements between the previous measurements is likely due the concentration uncertainty and the unknown initial xenon concentration in what is assumed to be pure liquid argon.

%% file: 6_LightYeildComparison.tex
\section{Light Yield Comparison}
The total increase in light yield resulting from Xe doping is evaluated with two methods. First, a direct comparison is made of the position of the muon peak seen in the pure Ar random trigger data to the muon peak in 10~ppm \XeD\ random trigger data. Second, the relative shift of the mean value of the random trigger distributions is measured. Cosmic muons can be seen in the random trigger data at large photo-electrons (PE) because the typical energies deposited are of order 100 MeV while nuclear decay energies are of order 1 MeV. The shape of the expected distributions were verified with an initial Geant4 simulation and from this simulation it is estimated that events with less than $\sim$100 PE are likely due to nuclear decays and not cosmic muons. The muon peak in both data sets is well modeled by a Landau function. Both data sets were fit in two different ranges in order to estimate this uncertainty. For the Ar data, the lower boundary of the fit range is fixed at 100~PE while the upper boundary is varied from 450 to 600~PE. A MPV of 334$\pm$21(syst)$\pm$7(stat)~PE is found. The fit for \XeD\ data had the lower bound fixed at 100~PE while the upper boundary is varied from 2000 to 10,000~PE. A MPV of 1004$\pm$47(syst)$\pm$4(stat)~PE is found. The position of this peak increased by a factor of 3.00$\pm$0.24(syst)$\pm$0.06(stat). If the mean of each distribution is calculated, the pure LAr data set has a mean value of 1067$\pm$68(syst)$\pm$4(stat)~PE while the 10~ppm \XeD\ data set has a mean value of 2050$\pm$20(syst)$\pm$14(stat)~PE which leads to a relative average light yield increase of 1.92$\pm$0.12(syst)$\pm$0.02(stat). This second method yields a smaller increase because a large portion of the photo-electron spectrum in the pure LAr spectrum falls above the the Landau peak and as such the spectrum as a whole is not well modeled by a single function. On the other hand, the distribution in the \XeD\ spectrum is fit well by a Landau. This light yield increase is larger than past results (\cite{CGWahl,Pieffer}) due to the longer attenuation length in \XeD\ at 175~nm in a larger active volume.  

\begin{figure}
\centering
\captionsetup{width=1\linewidth}
\includegraphics[width=0.49\linewidth]{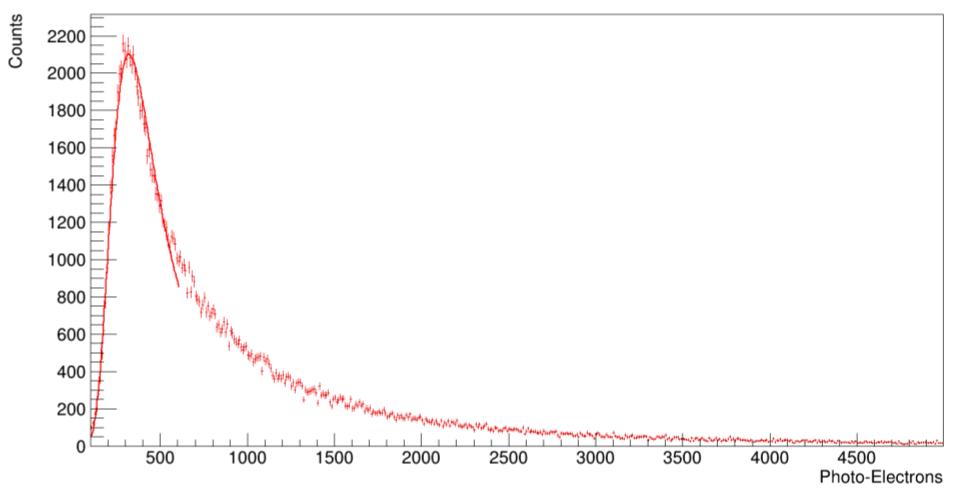}
\includegraphics[width=0.49\linewidth]{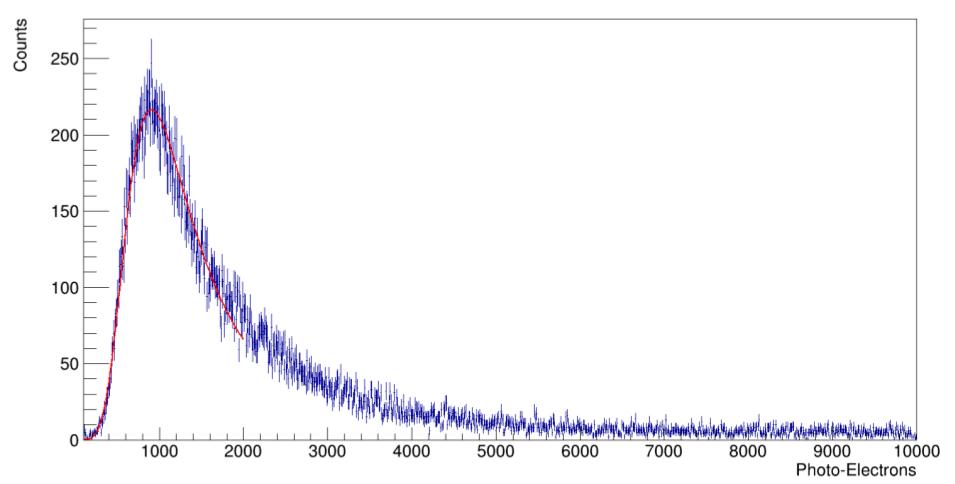}
\caption{(left) Muon peak from random trigger data in LAr. The solid red line is a Landau fit the peak. (Right) Muon peak from random trigger data in 10~ppm \XeD\ random trigger data.. The solid red line is a Landau fit the peak.}
\label{fig LightYeildIncrease}
\end{figure}

%% file: 7_conclusion.tex
\section{Conclusion}
\label{sec conclusion}

It has been demonstrated that a small concentration of gas can be reliably and precisely injected into a cryostat containing $\sim$120~kg of LAr. Nitrogen at a concentration of 1.06$\pm$0.07~ppm was inserted and the quenching factor (Birk's constant) was measured to be 0.12$\pm$0.03~ppm$^{-1}\mu$s$^{-1}$, consistent with past literature. This demonstrates that small quantities of gas can be precisely injected. The N$_2$ contaminant was then removed using a ullage recirculation and purification system with a filtration time constant as low as 3.6~d. 

We investigated \XeD\ concentrations of 1.00$\pm$0.06~ppm, 2.0$\pm$0.1~ppm, 5.0$\pm$0.3~ppm, and 10.0$\pm$0.5~ppm with the largest uncertainty being 6\% compared to past experiments which achieved 23.5\%. We have measured the xenon doping time constants as outlined in past experiments and give confidence to these measurements due to the small uncertainty in our doping concentrations. A relative light yield increase of 1.92$\pm$0.12(syst)$\pm$0.02(stat) is found for 10~ppm Xe compared to pure Ar.


%% file: 8_Acknowledgements.tex
\section*{Acknowledgements}
\label{sec Acknowledge}
We thank Stefan Sch\"{o}nert for useful conversations, Steven Boyd for his infinite wisdom concerning our cryogenics system, Dinesh Loomba for his help interpreting our data, and Matt Green and Brain Zhu for their expertise in Geant4. In view of the optical simulations, we acknowledge the development of the computationally efficient "heat-map" approach by the LEGEND simulations group (chaired by Matt Green) and the fruitful discussions we had with Luigi Pertoldi and Christoph Wiesinger from the GERDA collaboration. We would like to thank the Department of Physics and Astronomy at the University of New Mexico for the partial support of one of our authors (N.M.) and allowing us to move the experiment exceptionally early into the new physics building. This material is based upon work supported by the U.S.~Department of Energy, Office of Science, Office of Nuclear Physics under award number LANLE9BW. We gratefully acknowledge the support of the U.S.~Department of Energy through the LANL/LDRD Program for this work. Finally, we would like to acknowledge the U.S. Department of Energy, Office of Science, Office of Workforce Development for Teachers and Scientists, Office of Science Graduate Student Research (SCGSR) program. The SCGSR program is administered by the Oak Ridge Institute for Science and Education (ORISE) for the DOE. ORISE is managed by ORAU under contract number DE‐SC0014664. 

%% file: XeDopedArMaster.bbl
\begin{thebibliography}{10}

\bibitem{Aalbers:2016jon}
{\sc Aalbers, J., et~al.}
\newblock {DARWIN: towards the ultimate dark matter detector}.
\newblock {\em JCAP 1611\/} (2016), 017.

\bibitem{LEGENDOverview}
{\sc Abgrall, N., et~al.}
\newblock {The Large Enriched Germanium Experiment for Neutrinoless Double Beta
  Decay (LEGEND)}.
\newblock {\em AIP Conf. Proc. 1894}, 1 (2017), 020027.

\bibitem{DUNE}
{\sc Abi, B., et~al.}
\newblock {The DUNE Far Detector Interim Design Report Volume 1: Physics,
  Technology and Strategies}.
\newblock {\em arXiv\/} (2018).
\newblock 1807.10334.

\bibitem{Acciarri_2012}
{\sc Acciarri, R., Antonello, M., Boffelli, F., Cambiaghi, M., Canci, N.,
  Cavanna, F., Cocco, A.~G., Deniskina, N., Pompeo, F.~D., Fiorillo, G., and
  et~al.}
\newblock Demonstration and comparison of photomultiplier tubes at liquid argon
  temperature.
\newblock {\em Journal of Instrumentation 7}, 01 (Jan 2012), P01016–P01016.

\bibitem{triplet1.26}
{\sc Acciarri, R., et~al.}
\newblock {Effects of Nitrogen contamination in liquid Argon}.
\newblock {\em JINST 5\/} (2010), P06003.

\bibitem{Gerda}
{\sc Ackermann, K.~H., et~al.}
\newblock {The GERDA experiment for the search of $0\nu\beta\beta$ decay in
  $^{76}$Ge}.
\newblock {\em Eur. Phys. J. C73}, 3 (2013), 2330.

\bibitem{Agostini:2015boa}
{\sc Agostini, M., et~al.}
\newblock {LArGe: active background suppression using argon scintillation for
  the Gerda $0\nu \beta \beta $ -experiment}.
\newblock {\em Eur. Phys. J. C75}, 10 (2015), 506.

\bibitem{Agostini:2018}
{\sc Agostini, M., et~al.}
\newblock {Improved limit on neutrinoless double beta decay of $^{76}$Ge from
  GERDA Phase II}.
\newblock {\em Phys. Rev. Lett. 120\/} (2018), 132503.

\bibitem{xenonEmission}
{\sc Akerib, D.~S., et~al.}
\newblock {Liquid xenon scintillation measurements and pulse shape
  discrimination in the LUX dark matter detector}.
\newblock {\em Phys. Rev. D97}, 11 (2018), 112002.

\bibitem{xenonDopedPulseShape1}
{\sc Akimov, D., Belov, V., Konovalov, A., Kumpan, A., Razuvaeva, O., Rudik,
  D., and Simakov, G.}
\newblock {Fast component re-emission in Xe-doped liquid argon}.
\newblock {\em arXiv\/} (2019).
\newblock 1906.00836.

\bibitem{afterpulse:2017twm}
{\sc Butcher, A., Doria, L., Monroe, J., Retière, F., Smith, B., and Walding,
  J.}
\newblock {A method for characterizing after-pulsing and dark noise of PMTs and
  SiPMs}.
\newblock {\em Nucl. Instrum. Meth. A875\/} (2017), 87--91.

\bibitem{ArgonNitrogenXenon}
{\sc Buzulutskov, A.}
\newblock Photon emission and atomic collision processes in two-phase argon
  doped with xenon and nitrogen.
\newblock {\em arXiv 117}, 3 (Feb 2017), 39002.
\newblock 1702.03612.

\bibitem{LNGD_LE}
{\sc Chepel, V., and Araujo, H.}
\newblock {Liquid noble gas detectors for low energy particle physics}.
\newblock {\em JINST 8\/} (2013), R04001.

\bibitem{coldhead}
{\sc Cryomech}.
\newblock Al60 with cp820 cryorefrigerator specification sheet.
\newblock Material Reference, December 2007.

\bibitem{fields2020kinetic}
{\sc Fields, D.~E., Gibbons, R., Gold, M., Thomas, J.~L., McFadden, N.,
  Elliott, S.~R., Massarczyk, R., and Rielage, K.}
\newblock A kinetic model for xenon-doped liquid argon scintillation light,
  2020.

\bibitem{tpb}
{\sc Gehman, V.~M., Seibert, S.~R., Rielage, K., Hime, A., Sun, Y., Mei, D.~M.,
  Maassen, J., and Moore, D.}
\newblock {Fluorescence Efficiency and Visible Re-emission Spectrum of
  Tetraphenyl Butadiene Films at Extreme Ultraviolet Wavelengths}.
\newblock {\em Nucl. Instrum. Meth. A654\/} (2011), 116--121.

\bibitem{getter}
{\sc Getter, S.}
\newblock Ps4-mt3/15-r/n specifications specification sheet.
\newblock Material Reference, May 2002.

\bibitem{Himi:1982hf}
{\sc Himi, S., Takahashi, T., Ruan, J.~z., and Kubota, S.}
\newblock {LIQUID AND SOLID ARGON, AND NITROGEN DOPED LIQUID AND SOLID ARGON
  SCINTILLATORS}.
\newblock {\em Nucl. Instrum. Meth. 203\/} (1982), 153--157.

\bibitem{triplet1.6}
{\sc Hitachi, A., Takahashi, T., Funayama, N., Masuda, K., Kikuchi, J., and
  Doke, T.}
\newblock Effect of ionization density on the time dependence of luminescence
  from liquid argon and xenon.
\newblock {\em Phys. Rev. B 27\/} (May 1983), 5279--5285.

\bibitem{nobleGasConcentrations}
{\sc Hwang, S.-C. R. D. L. D. A.~M.}
\newblock {\em Noble Gases,Kirk-Othmer Encyclopedia of Chemical Technology (5th
  ed.)}.
\newblock Wiley, 2005.

\bibitem{flowMeter}
{\sc Instruments, S.}
\newblock Smarttrek100 data sheet.
\newblock Material Reference, June 2015.

\bibitem{ISHIDA1997380}
{\sc Ishida, N., Chen, M., Doke, T., Hasuike, K., Hitachi, A., Gaudreau, M.,
  Kase, M., Kawada, Y., Kikuchi, J., Komiyama, T., Kuwahara, K., Masuda, K.,
  Okada, H., Qu, Y., Suzuki, M., and Takahashi, T.}
\newblock Attenuation length measurements of scintillation light in liquid rare
  gases and their mixtures using an improved reflection suppresser.
\newblock {\em Nuclear Instruments and Methods in Physics Research Section A:
  Accelerators, Spectrometers, Detectors and Associated Equipment 384}, 2
  (1997), 380 -- 386.

\bibitem{triplet1.46}
{\sc Lippincott, W.~H., Coakley, K.~J., Gastler, D., Hime, A., Kearns, E.,
  McKinsey, D.~N., Nikkel, J.~A., and Stonehill, L.~C.}
\newblock Scintillation time dependence and pulse shape discrimination in
  liquid argon.
\newblock {\em Phys. Rev. C 78\/} (Sep 2008), 035801.

\bibitem{Neumeier_2015}
{\sc Neumeier, A., Dandl, T., Heindl, T., Himpsl, A., Oberauer, L., Potzel, W.,
  Roth, S., Schönert, S., Wieser, J., and Ulrich, A.}
\newblock Intense vacuum ultraviolet and infrared scintillation of liquid ar-xe
  mixtures.
\newblock {\em {EPL} (Europhysics Letters) 109}, 1 (jan 2015), 12001.

\bibitem{NeumeierLArAtten_2015}
{\sc Neumeier, A., Dandl, T., Himpsl, A., Hofmann, M., Oberauer, L., Potzel,
  W., Schönert, S., and Ulrich, A.}
\newblock Attenuation measurements of vacuum ultraviolet light in liquid argon
  revisited.
\newblock {\em Nuclear Instruments and Methods in Physics Research Section A:
  Accelerators, Spectrometers, Detectors and Associated Equipment 800\/} (Nov
  2015), 70–81.

\bibitem{Neumeier}
{\sc Neumeier, A., Dandl, T., Himpsl, A., Oberauer, L., Potzel, W., Schönert,
  S., and Ulrich, A.}
\newblock Attenuation of vacuum ultraviolet light in pure and xenon-doped
  liquid argon {\textemdash}an approach to an assignment of the near-infrared
  emission from the mixture.
\newblock {\em {EPL} (Europhysics Letters) 111}, 1 (July 2015), 12001.

\bibitem{Pieffer}
{\sc Peiffer, P., Pollmann, T., Schönert, S., Smolnikov, A., and Vasiliev, S.}
\newblock Pulse shape analysis of scintillation signals from pure and
  xenon-doped liquid argon for radioactive background identification.
\newblock {\em Journal of Instrumentation 3}, 08 (August 2008), P08007--P08007.

\bibitem{pmtR11065}
{\sc Photonics, H.}
\newblock R11065 preliminary data sheet.
\newblock Material Reference, May 2009.

\bibitem{G10}
{\sc Plastics, L.}
\newblock G10 technical data sheet.
\newblock Material Reference, June 2010.

\bibitem{phillipsScientific}
{\sc Scientific, P.}
\newblock Octal variable gain amplifier nim model 777 data sheet.
\newblock Material Reference, October 2019.

\bibitem{Zhijing}
{\sc Tang, Z.}
\newblock Simulation of filling tank with argon.
\newblock Fermi National Accelerator Laboratory Engineering Analysis, April
  2016.

\bibitem{CGWahl}
{\sc Wahl, C.~G., Bernard, E.~P., Lippincott, W.~H., Nikkel, J.~A., Shin, Y.,
  and McKinsey, D.~N.}
\newblock Pulse-shape discrimination and energy resolution of a liquid-argon
  scintillator with xenon doping.
\newblock {\em Journal of Instrumentation 9}, 06 (June 2014), P06013--P06013.

\bibitem{HWATT}
{\sc WATT, H.}
\newblock Diffusion in multicomponent gaseous mixtures, part 2. diffusion of
  xenon-133 in binary mixtures of xenon with helium, neon, argon, and krypton.
\newblock {\em Canadian Journal of Chemistry 43\/} (September 1964).

\bibitem{Zugec:2016}
{\sc Žugec, P., et~al.}
\newblock {Pulse processing routines for neutron time-of-flight data}.
\newblock {\em Nucl. Instrum. Meth. A812\/} (2016), 134--144.

\end{thebibliography}
